# Structure of a model TiO$_2$ photocatalytic interface

H. Hussain[1,2]†, G. Tocci[1], T. Woolcot[1], X. Torrelles[3], C. L. Pang[1], D. S. Humphrey[1], C. M. Yim[1], D. C. Grinter[1]§, G. Cabailh[4,5], O. Bikondoa[6], R. Lindsay[7], J. Zegenhagen[2]‡, A. Michaelides[1], G. Thornton[1]*

[1] London Centre for Nanotechnology and Department of Chemistry, University College London, 20 Gordon Street, London WC1H OAJ, UK.

[2] ESRF, 6 rue Jules Horowitz, F-38000 Grenoble cedex, France.

[3] Institut de Ciència de Materials de Barcelona (CSIC), Campus UAB, 08193 Bellaterra, Spain.

[4] Sorbonne Universités, UPMC Univ Paris 06, CNRS, 75005 Paris, France.

[5] CNRS, UMR 7588, Institut des NanoSciences de Paris, 75005 Paris, France.

[6] Department of Physics, University of Warwick, Gibbet Hill Road, Coventry C4 7AL, UK.

[7] Corrosion and Protection Centre, School of Materials, The University of Manchester, Sackville Street, Manchester, M13 9PL, UK.

* To whom correspondence should be addressed. Email: g.thornton@ucl.ac.uk

Current Addresses:

† Corrosion and Protection Centre, School of Materials, The University of Manchester, Sackville Street, Manchester, M13 9PL, UK.

§ Chemistry Department, Building 555, Brookhaven National Laboratory, Upton, NY 11973, USA.

‡ Diamond Light Source Ltd., Diamond House, Harwell Science and Innovation Campus, Didcot, Oxfordshire, OX11 0DE, UK.

**Abstract**

The interaction of water with $TiO_2$ is crucial to many of its practical applications, including photocatalytic water splitting. Following the first demonstration of this phenomenon 40 years ago there have been numerous studies of the rutile single crystal $TiO_2(110)$ interface with water. This has provided an atomic level understanding of the water/$TiO_2$ interaction. However, nearly all the previous studies of water/$TiO_2$ interfaces involve water in the vapour phase. In this Article we explore the interfacial structure between liquid water and a rutile $TiO_2(110)$ surface pre-characterised at the atomic level. Scanning tunneling microscopy and surface X-ray diffraction are used to determine the structure, which is comprised of an ordered array of hydroxyl molecules with molecular water in the second layer. Static and dynamic density functional theory calculations suggest that a possible mechanism for formation of the hydroxyl overlayer involves the mixed adsorption of $O_2$ and $H_2O$ on a partially defected surface. The quantitative structural properties derived here provide a basis with which to explore the atomistic properties and hence mechanisms involved in $TiO_2$ photocatalysis.

The generally accepted mechanism of photocatalysis by $TiO_2$ involves photoexcitation of electrons from the valence band to conduction band by light with energy greater than the 3 eV bandgap[1,2]. Holes in the valence band and electrons in the conduction band created by this excitation travel to the surface where they initiate chemical reactions. For example, the electrons can reduce water to hydrogen. The potential for harvesting light in this way to produce a portable fuel in the form of $H_2$ has motivated the study of technical catalysts. It has also motivated the study of model systems in the form of single crystal surfaces as a route to understanding the nature of the surface reactions at the atomic and molecular scale. Indeed, the interaction of water with $TiO_2$ in ultrahigh vacuum (UHV) has been extensively studied (see e.g. refs. 1,3–10). This is especially true for the most commonly explored rutile $TiO_2(110)$ surface (depicted in Fig. 1a), which is the lowest energy termination of rutile[11] and hence is the most appropriate model system for a technical catalyst.

The surface chemistry of water interacting with $TiO_2(110)$ under UHV is complex and has been the subject of considerable debate, mostly focusing on the level of water dissociation and the role of surface defects (see e.g. refs. 1,3–7,10,12). However, many aspects of the adsorption process have now been established. It is known, for example, that at room temperature water dissociates at bridging oxygen vacancies ($O_b$-vac) as well as <111> oriented steps, producing bridging OH ($OH_b$) groups[1,3–7,10,12]. These groups can be converted into terminal OH groups bound to 5-fold coordinated Ti atoms by reaction with $O_2$[13].

Although in the past, the emphasis has been on studies of the gas phase $H_2O$ interface with $TiO_2$, it is clear that the liquid/solid interface is more relevant for practical applications. Surface X-ray diffraction (SXRD) provides a potential means of elucidating the structure of this model photocatalytic interface at a quantitative level. This technique has been used extensively to determine metal/liquid interfacial structures under electrochemical control[14], which allows the metal surface to be cleaned *in situ*. This procedure is less straightforward for a semiconducting oxide substrate such as $TiO_2$. There has been a ground-breaking SXRD measurement of a $TiO_2(110)$-water interface[15], and one of the first near ambient pressure photoemission measurements investigated the chemical states at the interface between $TiO_2(110)$ and an ultrathin film of water[16]. However, in both cases a non-standard surface preparation method was employed. As for related modeling of the interface, there have been several computational studies of the water interface formed by the perfect $TiO_2(110)$ surface. The results of these calculations are controversial, being centred around the question of dissociation on the pristine surface[17–22].

Here we employ a novel approach to provide the first quantitative structure of a well-defined metal oxide-water interface, which also represents a model of the interface present in the rutile $TiO_2$ photocatalyst. More specifically, we perform both *ex situ* and *in situ* measurements of the liquid water-$TiO_2$ interface in an aerobic environment, formed by either temporarily immersing (dipping) a rutile $TiO_2(110)$ surface into water or by depositing a water droplet, respectively. Here we simplify the model photocatalyst to its oxide component in the absence of band gap light and metal co-catalyst. However, we note that UV light does not modify $TiO_2(110)$[23]. Moreover, the most effective co-catalyst is well-dispersed Pt nanoparticles[24], which are not expected to affect the $TiO_2(110)$-water component. Understanding a simple model system like the one considered here is an essential first step towards the characterisation of more complex $TiO_2$ photocatalysts.

The surfaces were characterised before and after exposure to liquid water using UHV STM and SXRD. STM measurements in conjunction with photoelectron spectroscopy in the same instrument evidence the formation of an ordered 2×1 hydroxyl overlayer formed after dipping. The SXRD results identify the bonding site as the 5-fold coordinated Ti atoms. This site is also occupied at the *in situ* liquid

water-$TiO_2$(110) interface, with ordering of molecular water in the second layer. This result is surprising based on what is known from UHV studies. It appears to arise from the availability of a small concentration of $O_2$ during the formation of the interface, which was not previously anticipated. This work demonstrates the importance of *in situ* structural characterisation of model photocatalytic interfaces and provides the basis for more realistic modeling of the photocatalytic interface with computational approaches.

In order for our liquid phase experiments to be connected with the results from $TiO_2$(110) UHV studies and to ensure accurate comparison with calculations[19,20,25], we employ UHV preparation and analysis methods that are known to produce and verify the presence of an atomically ordered substrate[1]. A section of the vacuum chamber (base pressures ~$1\times10^{-10}$ mbar) is then vented to $N_2$ (BOC, 99.998% purity) before dipping the sample in water or depositing a droplet of water to form a meniscus. The $N_2$ gas used has a nominal $O_2$ content of 5 ppm by volume, which introduces an aerobic environment as found in a real photocatalytic system. At the near-atmospheric pressure used for venting, this equates to a partial $O_2$ pressure of ~$5\times10^{-3}$ mbar and an equivalent exposure of ~$10^5$ Langmuir (1 L = $1.33 \times 10^{-6}$ mbar.sec).

Figure 1b shows STM images of the as-prepared $TiO_2$(110) surface[10]. In the high-resolution image of the inset, bright (Ti) rows are seen that run in the [001] direction. Bright spots are also present and these are a mixture of $O_b$-vacs and bridging hydroxyls ($OH_b$), the latter being formed by water dissociation at defect sites. After venting to $N_2$ and immersion in 10 ml water (18.2 MΩ.cm, total organic content < 2 ppm) for 5 mins, the sample ($H_2O_{dip}$ sample) was reintroduced to UHV. STM images of the $H_2O_{dip}$ sample are shown in Fig. 1c. The basic morphology of the surface is the same as that in Fig. 1b. There is no evidence of any pitting or erosion of the step edges following immersion in water[26].

The high-resolution image shown in the inset of Fig. 1c evidences a (2×1) overlayer. Antiphase domains of this overlayer can form by an offset of one unit cell along the [001] direction. The domains are small, most being shorter than ten $TiO_2$(110) unit cell lengths in [001] (~30 Å) and four units in [1$\bar{1}$0] (~26 Å). Half order low energy electron diffraction (LEED) beams were not observed, most likely due to either the small domain size and/or electron-stimulated desorption of the adsorbate. X-ray photoelectron spectroscopy (XPS) measurements indicate the level of C contamination to be around 0.1 monolayer (ML) (see Supplementary Fig. S1), where 1 ML is the density of primitive surface unit cells. Venting an as-prepared $TiO_2$(110) sample to air or pure $O_2$ without immersion in water was found not to form a (2×1) overlayer. A previous study also noted the absence of an ordered overlayer after exposure to a nitrogen atmosphere[27]. Hence, we conclude that the (2×1) overlayer is formed specifically by immersion in liquid water.

From the STM images, we find the coverage of the ordered overlayer to be 0.30±0.05 ML. This coverage and the domain size did not vary when we used samples with different initial $O_b$-vac concentrations of 0.16 and 0.07 ML. This suggests that neither $O_b$-vacs nor $OH_b$ (the coverage of which is proportional to the initial $O_b$-vac density) play a key role in the nucleation process. Hence, strain is a likely origin of the limited domain size[28].

To probe the chemical nature of the (2×1) overlayer, we employed XPS and UV photoelectron spectroscopy (UPS). A peak at ~532 eV binding energy in the O 1s spectrum that originates from OH becomes more intense after immersion.

Supplementary Fig. S2a,b shows a pair of XPS spectra taken before and after immersion of $TiO_2$(110) in water. Similarly, in UPS spectra taken after immersion in water, peaks appear at 8.0 and 10.2 eV below the Fermi level ($E_F$) that are characteristic of chemisorbed OH (Ref. 2) (Supplementary Fig. S2c). UPS also detects a band-gap state (BGS) associated with $O_b$-vac and $OH_b$ that lies ~0.8 eV below $E_F$[29–31]. This BGS is quenched after immersion in water, which can be explained by healing of the $O_b$-vacs/$OH_b$ by exposure to $O_2$ during the $N_2$ venting procedure (Supplementary Fig. S2d). The $O_2$ exposure of ~$10^5$ L is at least an order of magnitude larger than that required to attenuate the BGS (~400 L)[3,29].

Figure 2a shows selected SXRD results in the form of crystal truncation rods (CTRs) recorded from the $H_2O_{dip}$ surface together with those from the as-prepared UHV surface. The latter results are consistent with those reported previously[32]. Best fits to the data are also shown in Fig. 2a, with a larger dataset being presented in Supplementary Fig. S3. A model for the clean surface is shown in Fig. 2b and the best-fit model for the $H_2O_{dip}$ surface is shown in Fig. 2c. Supplementary Tables S1 and S2 show the atomic displacements, the key for each atom being obtained from Supplementary Fig. S4.

Due to its low X-ray scattering contribution, hydrogen is not included in the analysis and is not shown in the model. Instead, water molecules or hydroxyls are represented only by their oxygen atoms. The bond distance between O($1^\#$) and Ti(2) (see Fig. 2c) is 1.95 ± 0.03 Å, which is in good agreement with the literature value for the Ti–$OH_t$ (terminal OH) bond of 1.85 ± 0.08 Å[33], whereas the Ti–$H_2O$ bond is much longer at 2.21 ± 0.02 Å[34]. Hence, we assign O($1^\#$) to $OH_t$, which is also consistent with our photoelectron spectroscopy data. The SXRD data cannot rule out the presence of $OH_b$, however both the XPS and STM results from the $H_2O_{dip}$ surface suggest that only one form of OH species is present. In the case of XPS, the OH related peak in the O 1s spectrum increases by a factor of 5±1 after dipping. The STM image appears to contain only one type of adsorbate in the 2×1 overlayer.

As STM showed very little change to the morphology of the surface after immersion in liquid water, a near-perfect model was simulated, i.e. the occupancies were fixed to 1 during the fit, except that of the $OH_t$ molecule. This has an occupancy of 0.45±0.1, consistent with the STM results. If the occupancy of the $OH_t$ molecule is fixed at unity, $\chi^2$, the goodness of fit worsens to 1.6 from a minimum value 1.4 at 0.45 ML. This occupancy of below 0.5 is what one would expect for the (2×1) overlayer with domain wall absences and therefore supports the model. There is no intensity in the position of half-order rods, which is consistent with the small domain sizes found in STM.

The dipping (*ex situ*) experiments allow us to gain chemical composition as well as direct (STM) and reciprocal space (SXRD) information about the interface initially formed at the liquid water interface. Taken together, these measurements paint a robust picture of the interface. Only *in situ* SXRD measurements are possible at the liquid water interface ($H_2O_{drop}$ sample). As we shall show below, a clear connection can be made between the structures obtained from the *in situ* and *ex situ* measurements.

Selected CTRs of the $H_2O_{drop}$ surface are shown together with those from the $H_2O_{dip}$ and as-prepared UHV surfaces in Fig. 2. A larger data set is shown in Supplementary Fig. S5. The best-fit model (hydrated model) is shown in Fig. 2d and has a $\chi^2$ of 1.7. The model is essentially the same as that of the $OH_t$ model (Fig. 2c), except for the presence of a hydration layer above the (2×1) $OH_t$ contact layer. As with the $H_2O_{dip}$ sample, evidence for $OH_t$ (and not $H_2O$ molecules) at the $Ti_{5c}$ sites comes from the bond distance between $OH_t$ and Ti(2) of 1.95 ± 0.03 Å (see Supplementary Fig. S4 and Table S1). Our model differs from that derived from a previous SXRD measurement of a $TiO_2$(110)/liquid water interface. In this case an experimental Ti-$OH_t$ bond distance of 2.12 ± 0.02Å was found[15], which is closer to the Ti-$OH_2$ bond distance[34]. Our model also differs from that obtained from the near ambient pressure photoemission study, which concluded that bridging hydroxyls only were present at the interface[16]. In both cases the discrepancy will arise from the difference in sample preparation and characterisation. A 2×1 overlayer was recently observed in STM measurements of a $TiO_2$(110)-liquid water interface, where the substrate was prepared in the same manner employed here. By comparison with DFT calculations, water dimers are thought to form the overlayer[35]. However, on the basis of the work presented here, it seems more likely that the 2×1 overlayer arises from $OH_t$ groups.

To understand the formation of $OH_t$ from exposure of the $TiO_2$(110) surface to liquid water, we note that UHV-prepared samples contain $O_b$-vac and that these react with water in the residual vacuum to form $OH_b$ (ref. 3–7,10,12). Approximately $10^5$ L $O_2$ is also supplied here during the $N_2$ venting procedure. Thus water, $O_2$(g), $OH_b$, and $O_b$-vac are all potential reactants when we expose our $TiO_2$(110) surfaces to liquid water and when combined, $OH_t$ is a likely candidate product[13]. Subsurface defects, such as sub-surface Ti-interstitials ($Ti_{int}$), can also participate in the reaction process, by providing a source of excess electrons for the dissociation of $O_2$(g)[30,36].

With the above considerations in mind, we performed an extensive series of density functional theory (DFT) calculations to understand the formation of the (2 × 1) $OH_t$ overlayer. Both geometry optimisations for the interface under UHV-like conditions and *ab initio* molecular dynamics (AIMD) under aqueous conditions were performed. Taken together these suggest that this overlayer most likely forms through the mixed dissociation of $O_2$ and $H_2O$ on a $TiO_2(110)$ surface containing point defects, in the form of $OH_b$, oxygen adatoms ($O_{ad}$) and $Ti_{int}$. Specifically, in Fig. 3 we show the potential energy diagram obtained for the vacuum interface for a possible mechanism through which the (2 × 1) $OH_t$ overlayer can form. The mechanism that governs the stabilisation of the overlayer relies on a competition between charge transfer arising from the presence of defects (see Supplementary Fig. S6) and the surface distortion due to the adsorption of the $OH_t$[37,38].

In the process shown in Fig. 3, the energy zero reference state (see Fig. 3(a),(b)) is represented by a $TiO_2(110)$ surface model with 1/4 ML of $OH_b$, 1/8 ML of $O_{ad}$ and 1/4 ML of subsurface $Ti_{int}$. From this initial state the formation of the (2 × 1) $OH_t$ overlayer proceeds with the adsorption of $O_2$ from the gas phase on a $Ti_{5c}$ site adjacent to a $OH_b$ site (Fig. 3(c)). $O_2$ can then react with the $OH_b$ to form a pair of $OH_t$ (Fig. 3(d)). Following this step, water may adsorb on $Ti_{5c}$ sites to form a mixed overlayer made of an $OH_t$ pair, an $O_{ad}$ and three $H_2O$ (Fig. 3(e)). Through a sequence of proton transfer events involving the water and the $OH_t$ pair, as well as water and the $O_{ad}$, a (2 × 1) $OH_t$ overlayer can form. In this state (see Fig. 3(f)) two water molecules are co-adsorbed with four $OH_t$, with four $OH_t$ arranged in a (2 × 1) symmetry. The $4OH_t+2H_2O$ and the $2OH_t+O_{ad}+3H_2O$ states are almost degenerate (the $4OH_t+2H_2O$ state is about 50 meV less stable). In the final step shown in Fig. 3 water is desorbed to the gas-phase, leaving only a (2 × 1) overlayer of $OH_t$ (Fig. 3(g)). From Fig. 3 it is clear that the bare $OH_t$ (2 × 1) overlayer is less stable than the state where water is present. This result is consistent with our STM measurements, where it has been observed that the $OH_t$ species are arranged in (2 × 1) symmetry only after dipping the sample in water. Under aqueous conditions instead, the $OH_t$ (2 × 1) overlayer may be stabilised by the presence of water molecules from the liquid phase. This is indeed what we find from analysis of our AIMD simulations of the interface under aqueous conditions, where we find that: (i) the $OH_t$ (2 × 1) overlayer remains stable over two separate AIMD simulations with a total length of 70 ps; and (ii) on average 0.18 ML of the $Ti_{5c}$ sites adjacent to the terminal OH groups are occupied by water molecules.

The AIMD simulations also provide an understanding of the second layer water structure in the case of the $H_2O_{drop}$ surface. Figure 4 shows a comparison between the results from SXRD and from AIMD on the structure of the top $TiO_2$ layers and of the first two overlayers. Figure 4a illustrates a typical structure extracted from an AIMD trajectory of a liquid water film on $TiO_2(110)$ with 1/4 ML $Ti_{int}$ and the (2 × 1) $OH_t$ overlayer structure. The histograms in Fig. 4b show a comparison between the number of O and Ti atoms per unit cell obtained from SXRD and from AIMD as a function of the height from the surface. It can be seen that the histogram of the number of O and Ti atoms computed from AIMD overall agrees well with that extracted from SXRD. In terms of the location of the first adsorption layer, AIMD predicts a height of 2.07 Å, which is only moderately larger than the height of 1.95±0.03 Å measured in SXRD. The location of this first peak is actually rather similar to what one obtains from AIMD and classical force field simulations for both intact and dissociated water on $TiO_2$[15,19]. Our AIMD simulations predict that there are on average 0.68 O atoms per unit cell in this layer and also that a fraction of water (about 1/8 molecules per unit cell) diffuse in and out of this layer during the course of the simulation (see Supplementary Movie S1). The second adsorption layer consists of water molecules that are H-bonded to $OH_t$, to the transient molecular water in the first adsorption layer, or to $O_b$ atoms. Our AIMD simulations predict that this layer is at a height of 3.98 Å from the top surface layer, in reasonable agreement with the height extracted from SXRD of 3.80 ± 0.04 Å. It is interesting to note that our AIMD simulations do not show any evidence of facile proton transfer within the (2 × 1) $OH_t$ overlayer. We cannot exclude that proton hopping and/or proton diffusion events could occur on longer timescales than we have been able to simulate. Nevertheless, the rate for such processes at this interface would still be significantly lower than proton transfer events in liquid water or at other liquid water/oxide interfaces (see e.g. 39,40). As shown in the SI, individual proton transfer events invariably destabilise the overlayer by causing pairs of OH groups to be adsorbed at adjacent adsorption sites.

Our measurements were carried out at near neutral pH, which is the optimum for photocatalysis by $TiO_2$/Pt (see ref. 24). Thermodynamic models suggest that at pH 7 there could be a coverage of hydroxyls on $TiO_2$ associated with a pH greater than the point of zero charge[18]. The latter has been measured to be around 5 for the (110) termination[41]. To examine this further, we carried out an additional AIMD simulation to examine the possibility that the $OH_t$ overlayer was formed from the type of $OH^-$ diffusion to the oxide/water interface predicted by an AIMD study of liquid water on anatase $TiO_2(101)$[42]. The results of this simulation are shown in the Supplementary Fig. S7. An $OH_t$ (2 × 1) overlayer made of $OH^-$ ions is found to be unstable as evidenced by frequent proton transfer events, which disrupt the (2 × 1) symmetry and also result in the formation of O-adatoms. This suggests that the $OH_t$ (2 × 1) overlayer instead arises from a mixed $O_2$ and $H_2O$ dissociation in the presence of point defects as described above.

In summary, we have shown that a model for a photocatalytic interface between liquid water and rutile $TiO_2(110)$ has terminal hydroxyls in the contact layer. This picture comes from a combination of data from real space imaging, spectroscopic measurements and surface X-ray diffraction, with interpretation aided by DFT calculations. The ideal coverage of $OH_t$ is half a monolayer, corresponding to a (2 × 1) structure, but this is decreased to approximately 0.4 monolayers by absences at domain wall boundaries. This interface structure, created in the aqueous aerobic environment considered here, had not been anticipated and is different from what has been established for water on $TiO_2$(110) under UHV conditions. It is likely to have important implications for the chemistry of wet $TiO_2$ surfaces. For example, proton hopping and proton transfer is likely to be more facile at an aqueous interface with a high proportion of hydroxide groups (see e.g. ref [40]). Of greatest importance, perhaps, is that the energetics of each elementary step and possibly also the mechanism of the water oxidation process could be altered by the new interface structure identified here[43,44]. Understanding of water oxidation processes in $TiO_2$ interfaces under operating aqueous conditions is key to improve the activity of photoelectrochemical cells. Because the presence of $OH_t$ is likely to open up new reaction pathways to water oxidation, it could be possible to enhance the degree of hydroxylation further through e.g. self-doping of the substrate and exposure to $O_2$. We hope that by providing accurate experimental structural data for a model photocatalytic interface, this work paves the way to the investigation of the elementary steps involved in water splitting, including a treatment of the excited state electronic properties. Ultimately, this will lead to an atomistic level understanding of the photocatalytic process of water splitting in more complex systems.

**Methods**

The $TiO_2$(110) (Pi-Kem) samples were prepared following an established procedure[10] involving cycles of $Ar^+$ ion sputtering and annealing to ~1000 K in UHV. Auger electron spectroscopy (AES) or X-ray photoelectron spectroscopy (XPS) were used to confirm the surface cleanliness and low energy electron diffraction (LEED) or scanning tunneling microscopy (STM) were used to ensure that the surfaces were well-ordered and unreconstructed. For all three methods, any adventitious UV-induced modification was avoided by using a red light. Prior to dipping, ultra-pure water was de-aerated by bubbling $N_2$ through the liquid for >1 hour. This results in a pH of 6.9.

**Scanning tunneling microscopy (STM).** STM measurements were performed with an *Omicron* UHV variable temperature STM. Tunneling was into empty sample states with the sample bias ($V_s$ = 1.2-1.5 V) and tunneling current ($I_t$ = 0.2-0.5 nA). The sample was dipped in ultra pure, deoxygenated $H_2O$ for a range of volumes (2 ml - 2 l) and times (15 s – 10 min). Surface contamination was minimised by purging the load-lock with $N_2$. Immediately after the immersion, the sample was reintroduced into UHV where it was transferred to the analysis chamber for STM measurements.

**Photoelectron Spectroscopy.** Normal emission measurements were performed at room temperature in an *Omicron* UHV low temperature STM system, incorporating an *Omicron* HA125 hemispherical energy analyser. UV photoelectron spectra (UPS) used He II (hv = 40.80 eV) and He I (hv = 21.20 eV) excitation in order to monitor the valence band and band-gap regions. The Fermi energy ($E_F$) was determined from the tantalum sample holder that was in electrical contact with sample. Samples were also monitored using the *in situ* STM to ensure the presence of a 2×1 overlayer.

**Surface X-ray Diffraction (SXRD).** All SXRD measurements were carried out on the ID32 beamline at the European Synchrotron Radiation Facility (ESRF), employing UHV facilities located in the associated surface characterisation laboratory for sample preparation. Once prepared, the sample was transferred under UHV to one of two bespoke, small portable UHV chambers that were mounted on a six-circle diffractometer with the sample surface in the horizontal plane. For the $H_2O_{dip}$ measurements, data acquisition and analysis followed the procedures described in ref. 31 and therein. For $H_2O_{drop}$ measurements, a total of 1450 non-equivalent reflections were measured, employing an 'electrochemical cell' apparatus[45]. $H_2O_{dip}$ measurements, following a similar procedure to STM measurements, were conducted on a sample that was dipped in ~20 ml of ultra-pure, deoxygenated $H_2O$ for approximately 15 s. A large data set of 20 CTRs was collected that, after corrections, comprised of 835 non-equivalent reflections. The following reflections were monitored at regular intervals to ensure that there was no sample degradation: (1,0,1) $H_2O_{dip}$; (1,0,0.2) $UHV_{as\text{-}prepared}$; (1,3,0.05) $H_2O_{drop}$. Data for the latter reflection are shown in Fig. S8. The best-fit model takes into account Ti sites that are both occupied/unoccupied with OH, with the latter simulating domain walls.

**Theoretical.** Adsorption of the various species on $TiO_2(110)$ was modeled through spin-polarised DFT structure optimisations using periodic supercells ranging from (2×1) to (4×2). Adsorption on one side of a four tri-layer $TiO_2(110)$ surface was considered with the bottom tri-layer fixed at its bulk position and at least 15 Å separating periodic images along the surface normal. We considered a degree of hydroxylation of the surface ranging from 1/4 to 1 ML and a range of $Ti_{int}$ located in the second subsurface layer in a range of concentrations between 0 and 1/2 ML (See Supplementary Fig. 5). The reference states for the adsorption of $O_2$ or $H_2O$ (See Fig.

3 and Supplementary Fig. 5, for $O_2$ only) were taken to be their total energies in the gas phase. Interstitials were located in the second subsurface layer, and a very small dependence of the adsorption energy of $O_2$ on the position of the $Ti_{int}$ has been observed[37]. The VASP[46,47] code was used for these calculations with a plane wave cut-off of 400 eV and projector augmented wave potentials[48]. Sampling of the Brillouin zone was achieved using a Γ-centred (4×2×1) per primitive surface cell k-point mesh. The PBE[49] exchange-correlation functional with a Hubbard-U correction of 4.2 eV was used. By comparing our results with the hybrid functional HSE06[50] (see Supplementary Information) we ensured that the chosen value of U did not significantly affect the results regarding the formation of the (2×1) $OH_t$ overlayer.

The other set of simulations performed as part of this study involved spin-polarised DFT based molecular dynamics. These AIMD simulations were performed in order to investigate the structure of the liquid water/$TiO_2$ interface and the stability of the $OH_t$ (2×1) overlayer under aqueous conditions. The computational details are similar to those reported in our previous work (see refs 19,20) but are included here for clarity. In brief we used the CP2K/QUICKSTEP[51] code with the PBE exchange-correlation functional. CP2K/QUICKSTEP employs a mixed Gaussian and plane-wave basis set and norm conserving pseudopotentials. We used a short-range double valence polarised Gaussian basis[52] and a 480 Ry cut-off for the plane wave expansion. The AIMD trajectory analyzed in the main text is 50 ps in length, with a 1 fs timestep and deuterium masses for the hydrogens in the canonical ensemble with a target temperature of 360 K. As with the static DFT calculations, $TiO_2$(110) was modeled using a (4×2) unit cell, comprised of four tri-layers and 15 Å of vacuum. The liquid water film in contact with the surface was ca. 2 nm thick, being comprised of 87 molecules. The (2×1) overlayer was modeled with neutral $OH_t$ adsorbed at $Ti_{5c}$ sites with $Ti_{int}$ positioned in the second subsurface layer at a concentration of 1/4 ML. We also performed an additional 50 ps-long AIMD simulation without interstitials to test their effect on the stability of the $OH_t$ overlayer. The results of these tests show that the $OH_t$ (2×1) overlayer is only stable in the presence of interstitials and with neutral OH species as opposed to $OH^-$ (results are shown in the Supplementary Figs. S5,S6). Finally, we performed a number of additional AIMD simulations (each of which is about 20 ps-long) of $TiO_2$(110) under aqueous conditions as well as static DFT calculations under UHV-like conditions to compare the stability of the $OH_t$ (2×1) overlayer against that of other less ordered overlayers and to investigate the proclivity of proton transfer at the water/$TiO_2$(110) interface. Overall, we find that proton transfer and proton diffusion is much more facile at less ordered structures compared to the $OH_t$ (2×1) overlayer (see Supplementary Figs. S9-S11).

**Acknowledgments:**

The authors would like to thank Marco Nicotra, Yu Zhang and Michael Allan for assistance with some measurements. This work was funded by grants from the EPSRC (UK) (EP/C541898/1), M.E.C. (Spain) through project MAT2012-38213-C02-02, EU ITN SMALL, EU COST Action CM1104, ERC Advanced Grant (GT, ENERGYSURF No. 267768), ERC Consolidator Grant (AM, HeteroIce project No. 616121) and the Royal Society. We are grateful to the London Centre for Nanotechnology and UCL Research Computing for computation resources, and to the UKCP consortium (EP/ F036884/1) for access to Archer.

**Author Contributions**

G.Thornton, J.Z., A.M. designed the project. D.S.H., T.W., C.L.P., C.M.Y., D.C.G., H.H. performed the STM measurements and T.W. analysed the data.  T.W., C.M.Y. performed the UPS experiments with T.W. analysing the data. H.H., G.C., O.B., X.T., R.L., G.Thornton performed the SXRD measurements and H.H. and X.T. analysed the data. G.Tocci and A.M. conceived, designed and analysed the ab initio calculations. G. Tocci performed the ab initio calculations. H.H., T.W., G. Tocci, C.L.P., A.M., G.Thornton wrote the manuscript and the Supplementary Information with input from all authors.  All authors participated in discussing the data.

**Additional information**

The authors declare no competing financial interests. Supplementary Information accompanies this paper on www.nature.com/naturematerials. Reprint and permissions information is available online at http://npg.nature.com/reprintandpermissions. Correspondence and requests for materials should be addressed to G.Thornton.

**Figure Legends**

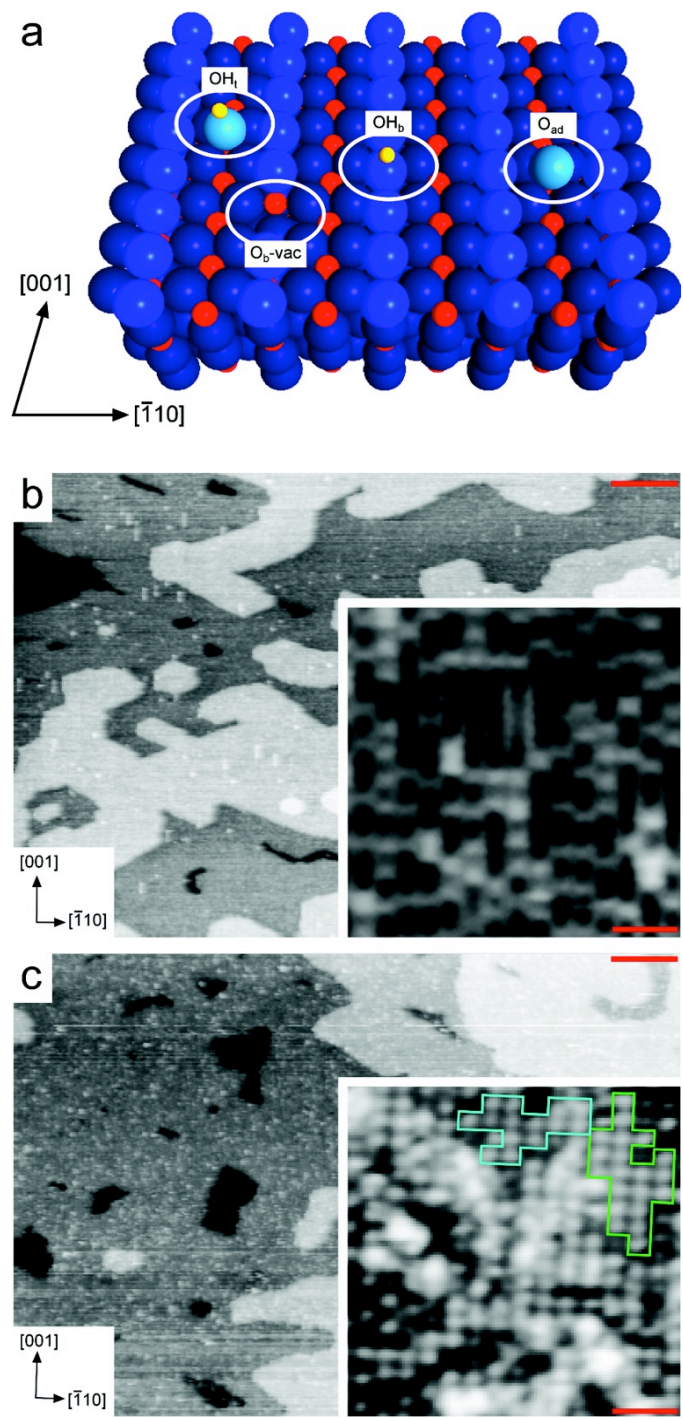

**Figure 1**. **The TiO$_2$(110) surface. a,** Ball model of TiO$_2$(110) together with O$_b$-vac, OH$_b$, OH$_t$, and O$_{ad}$. Ti is shown red and O blue: O$_b$ is shown lighter, and adsorbed O

lighter still. H is shown yellow. **b, c,** STM images (0.2 nA, Vs=1.2 V) of the surface before and after immersion in liquid water, respectively. In the large-area images, the red scale bars are 100 Å and in the insets 20 Å. Before immersion, the $TiO_2$(110) surface is characterised by bright rows that arise from $Ti_{5c}$ atoms and bright spots that correspond to $O_b$-vac (0.11 ML) and $OH_b$ (0.10 L). After immersion in liquid water to form the $H_2O_{dip}$ sample, the surface is characterised by bright spots arranged with a (2×1) periodicity. The blue and green shapes mark areas that are particularly well ordered. Note that these two areas are antiphase domains; the bright spots are offset by one $TiO_2$ surface unit cell in the [001] direction.

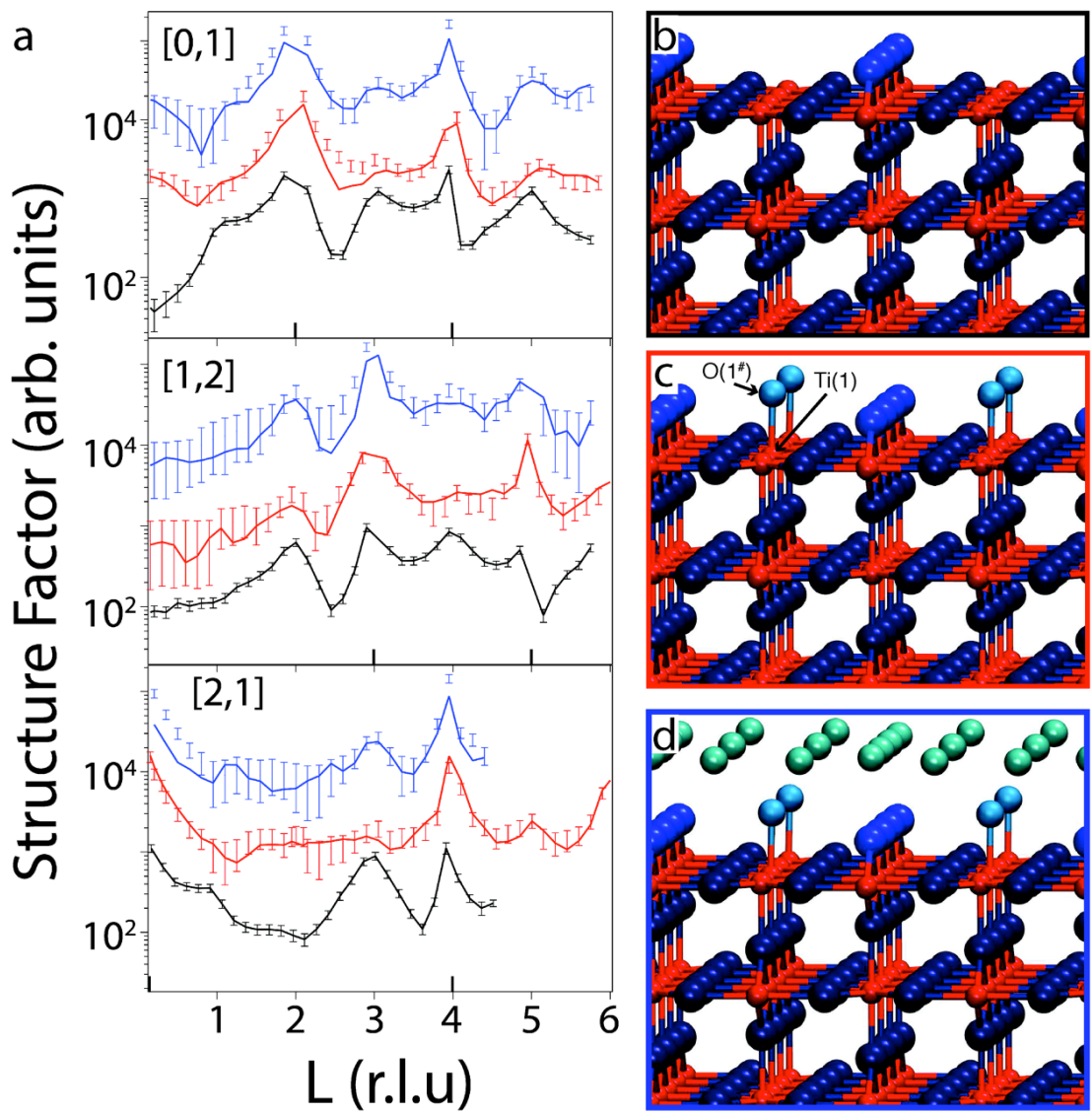

**Figure 2. Selected CTRs from the SXRD measurements alongside proposed models**. **a**, The structure factors of the different $TiO_2(110)$ surfaces are plotted for the (0,1,l), (1,2,l) and (2,1,l) CTRs. Black, red, and blue error bars represent the data from the as-prepared surface recorded before the $H_2O_{drop}$ experiment, $H_2O_{dip}$, and $H_2O_{drop}$ samples, respectively, with solid lines being the calculated data. Profiles are offset for clarity. Notches on the *x* axes correspond to Bragg peaks. **b-d**, Ball-stick models for clean, $H_2O_{dip}$, and $H_2O_{drop}$ samples, respectively.

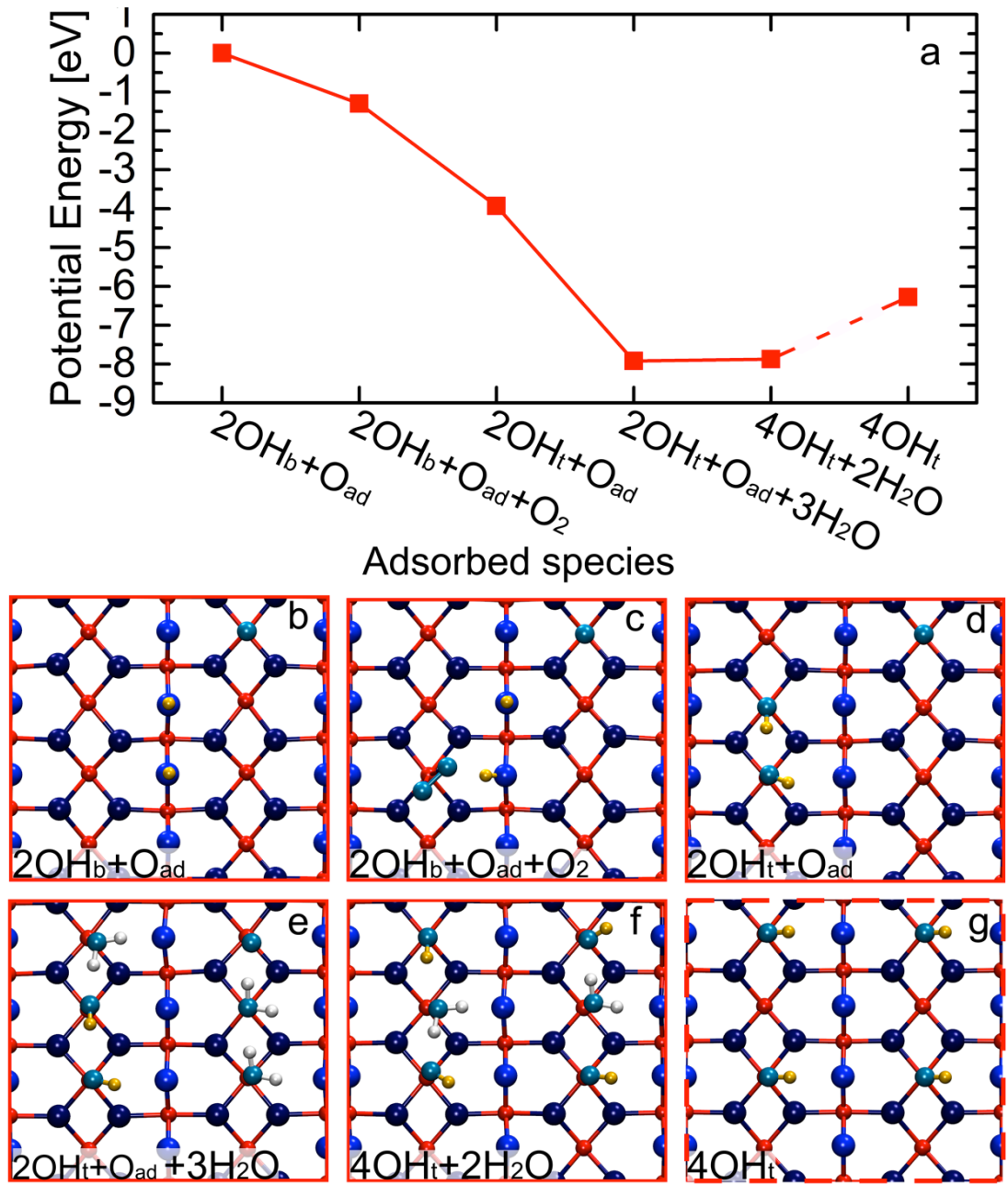

**Figure 3. Possible sequence of reaction steps leading to the formation of the OH$_t$ (2 × 1) overlayer. a,** Potential energy diagram for the formation of the OH$_t$ overlayer resulting from the mixed adsorption of O$_2$ and H$_2$O on a defective TiO$_2$(110) surface model. The value of the potential energy $\Delta E$ at each state $s$ is obtained as the difference between the total energy of that state $E_s$ and that of the previous state $E_{s-1}$, i.e. $\Delta E_s = E_s - E_{s-1} - E_s^{(gas-ads)} + E_s^{(gas-des)}$, and where $E_s^{(gas-ads)}$ and $E_s^{(gas-des)}$ are the total energy of any gas-phase specie that has been adsorbed or desorbed upon going from state $s-1$ to state $s$, respectively. The index $s$ goes from 1 to 5 and the state for $s = 0$ corresponds to the reference zero state composed of a surface with 1/4 ML of subsurface $Ti_{int}$, 1/4 ML of $OH_b$, and 1/8 ML of $O_{ad}$. The values of the potential energy refer to a (**4×2**) unit cell. The dashed line connecting the $4OH_t + 2H_2O$ state with the $4OH_t$ state indicates the desorption of 2 water molecules to the gas phase. **b-g,** Structures of adsorbates on the defective $TiO_2(110)$. The configurations are labeled according to the states shown in the potential energy diagram in **a**.

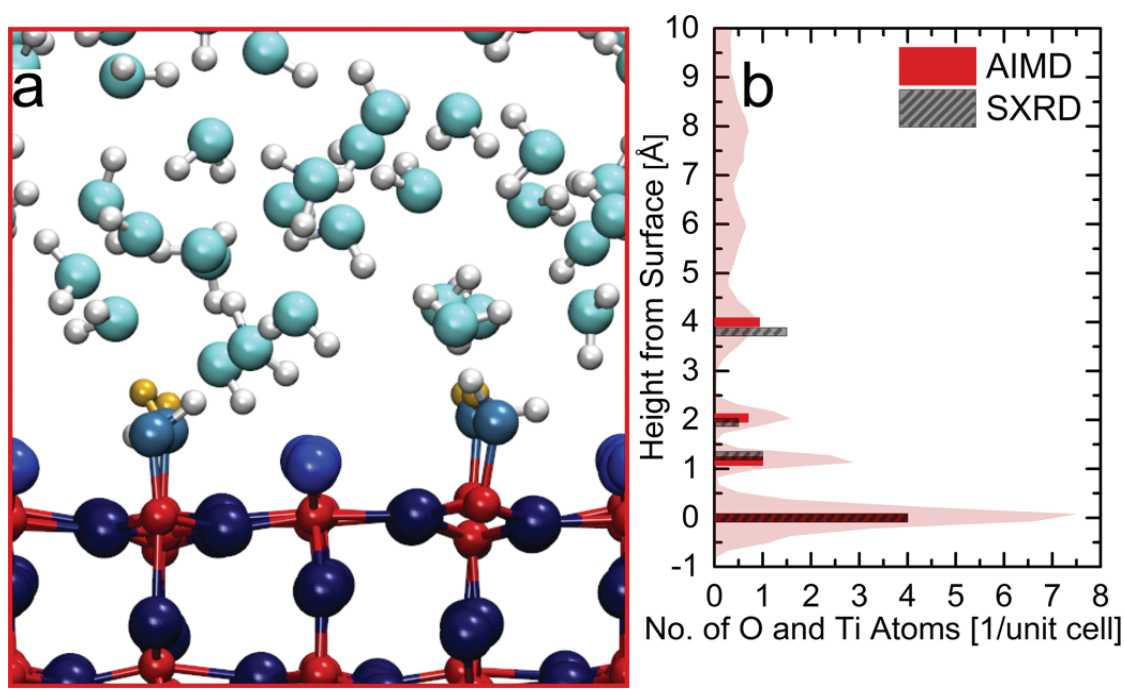

**Figure 4. Structure of the water/TiO₂ interface in aqueous conditions.** (a) Snapshot of a liquid water film on $TiO_2$ obtained from AIMD simulation. (b) Comparison between the number of Ti and O atoms per unit cell obtained from AIMD and from SXRD as a function of the height from the surface. The zero in the height corresponds to the top surface layer of the Ti$_{5c}$ atoms. The shaded red curve in b is the number density of the O and Ti atoms per unit cell $n(z)$ as a function of the height $z$, obtained from the AIMD. Integration of $n(z)$ between the minima of each peak $p$ (delimited by the heights $z_p$ and $z_{p+1}$) gives the number of O and Ti atoms per unit cell for a given layer $N(\bar{z}_p)$, which is shown in b by the red bar charts for each interface layer: $N(\bar{z}_p) = \int_{z_p}^{z_{p+1}} n(z)dz$, and where $\bar{z}_p = \int_{z_p}^{z_{p+1}} z\, n(z)dz / \int_{z_p}^{z_{p+1}} n(z)dz$.